\documentclass[%
 reprint,
 amsmath,amssymb,
 aps,
prb,
floatfix,
]{revtex4-2}

\usepackage{graphicx}
\usepackage{dcolumn}
\usepackage{bm}

\usepackage{refcount}

\usepackage{physics}
\usepackage[percent]{overpic}

\usepackage{hyperref}
\hypersetup{
colorlinks=true,
linkcolor=cyan,
filecolor=blue,
urlcolor=blue,
citecolor=green,
}
\newcommand{\refRomanLink}[1]{
    \hyperref[#1]{\Roman{\getrefnumber{#1}}}
}
\begin{document}

\preprint{APS/123-QED}


\title{Extrinsic nonlinear spin currents in spin-orbit coupled systems: a Boltzmann transport study}

\author{Ji-Wei Tian}
\affiliation{School of Physics, Peking University, Beijing 100871, China}

\author{H. Huang}
\affiliation{School of Physics, Peking University, Beijing 100871, China}

\date{\today}

\begin{abstract}
We develop a semiclassical Boltzmann framework for second-order extrinsic spin transport in the presence of skew scattering. The conventional relaxation-time approximation reduces the full collision integral to a single phenomenological timescale and thereby neglects the angular structure of impurity scattering. While the transport relaxation time primarily characterizes the leading angular harmonic of the nonequilibrium distribution, nonlinear response is sensitive to its higher angular harmonics, which are governed by distinct components of the collision operator. We therefore retain the angular and band dependence of the scattering kernel and decompose both the collision integral and the distribution function into Fourier harmonics. The resulting coupled algebraic equations allow the symmetric and skew-scattering corrections to the distribution function to be obtained recursively order by order in the electric field. We apply this formalism to the gapped surface states of a topological insulator with spin-dependent impurity scattering and evaluate the corresponding second-order extrinsic spin-current response. Our results show that higher angular harmonics of the nonequilibrium distribution and scattering between distinct constant-energy contours play an essential role in extrinsic nonlinear spin transport beyond the single-relaxation-time approximation.
\end{abstract}

\maketitle


\section{\label{sec:level1}Introduction \protect}
In recent years, the nonlinear response of electronic systems has garnered significant interest due to its potential applications in characterizing and measuring electronic and spintronic devices, as well as its relevance to fundamental physics. Among the various nonlinearresponse effects, second-order charge and spin Hall responses have attracted particular attention because they generate transverse charge or spin currents in response to a longitudinal electric field.\cite{PhysRevLett.115.216806, ma_observation_2019,zhang_quantum_2019, du_quantum_2021,PhysRevLett.123.196403, PhysRevLett.123.016801}
\par
Nonlinear transport phenomena can be categorized into intrinsic and extrinsic components, depending on their dependence on impurity scattering. The first systematically investigated extrinsic contribution is the Berry curvature dipole, which is proportional to the transport lifetime $\tau$ and relates the scattering to the geometry of the band structure. Unlike linear Hall conductivity, the Berry curvature dipole does not vanish in the presence of time reversal symmetry($\mathcal{T}$). Consequently, it could serve as a valuable tool for detecting the band geometry when $\mathcal{T}$ remains unbroken. Recent studies have also identified intrinsic second-order Hall responses associated with field-induced corrections to the Berry connection and the Berry-connection polarizability tensor(BCP). While some research has been conducted on the nonlinear spin Hall effect, these studies have either examined the nonlinear Drude component of the spin current or focused on systems without spin-orbit coupling (SOC), where the impact of spin is not decisive. Importantly, many existing semiclassical treatments employ a single relaxation-time approximation to address the extrinsic current, even though extrinsic mechanisms such as skew scattering and side jump are critical to the extrinsic spin Hall effect.\cite{RevModPhys.82.1539,PhysRevB.74.245309,PhysRevB.64.014416,PhysRevLett.92.126603,PhysRevLett.104.186403,vignale_ten_2010}\par
In this work, we develop a Boltzmann framework that retains the angular dependence of both symmetric and antisymmetric impurity scattering. By expanding the distribution function and the collision kernel in angular harmonics, the integral Boltzmann equation is reduced to a set of coupled algebraic equations for each harmonic. This construction allows the symmetric and skew-scattering corrections to the distribution function to be obtained recursively to second order in the electric field. We apply the method to the gapped surface states of a topological insulator and examine how the nonlinear spin-current response depends on the chemical potential, the band dispersion, and the available interband and inter-contour scattering channels.\par

The paper is structured as follows: In Section $\mathrm{\ref{sec:background}}$, we offer a concise review of spin-orbit coupling (SOC) and the origin of skew scattering by impurity atoms, accompanied by a T-matrix representation of skew scattering. We show the different effective relaxation time for the symmetric and skew part of the electron distribution function. In Section $\mathrm{\ref{sec:Method}}$, we present a method for calculating the electron distribution function order by order through the direct treatment of the collision integral and demonstrate its application using the surface states of a topological insulator as an exemplary model. Section $\mathrm{\ref{sec:result}}$ shows the results of the calculation of the exemplary model and offers analysis on these results. Finally, in Section $\mathrm{\ref{sec:scale}}$, we examine the scaling behaviors of nonlinear transport effects.

\section{\label{sec:background} SKEW SCATTERING}

\subsection{\label{sec:bg1} Spin-orbit Interaction}

Spin-orbit coupling (SOC)\cite{manchon_new_2015} plays an important role in spin transport phenomena, such as the spin Hall effect.
The general form of SOC is:
\begin{equation}
    H_{SOC}=\lambda_{SOC}(\vb{p}\times\grad V(r))\cdot\vb{\sigma} 
    \label{eq:1}
\end{equation}
where $\vb{p}$ is the momentum of electron and $V(r)$ is the potential acting on the electron, and $\vb{\sigma}$ are the $2\times2$ Pauli matrices of spin.\par
Equation (\ref{eq:1}) is our starting point to describe SOC. However, in practice, it is more convenient to construct an effective Hamiltonian in solids. In particular, in 2D systems, different SOC terms emerge in 2D systems, such as the intrinsic SOC, Rashba SOC, and Dresselhaus SOC.
\subsection{\label{sec:bg2} Single Impurity Skew Scattering}
Skew scattering is one of the extrinsic mechanisms that contribute to the spin Hall effect\cite{RevModPhys.87.1213}, and it is a well-known phenomenon that was initially studied by Mott to describe the scattering of spin-polarized electrons with atoms in the presence of SOC\cite{Mott1929}. The Hamiltonian of the single impurity scattering problem is given by:
\begin{equation}
    H = \frac{\vb{k}^2}{2m^*}+V_{imp}(\vb{r})+\lambda_{SOC}(\vb{p}\times V_{imp}(\vb{r}))\cdot \vb{\sigma}
    \label{eq:2}
\end{equation}
where $m^{*}$ is the effective mass of electrons, and $V_{imp}(\vb{r})$ is the potential of the impurity. The scattering amplitude between the initial state $\ket{\vb{k}\sigma}$, where $\vb{k}$ is the momentum of electrons and $\sigma$ is the $z$ component of spin, and the final state $\ket{\vb{k}^{\prime}\sigma^{\prime}}$, can be calculated based on the T-matrix method\cite{RevModPhys.82.1539}. Under the second-order Born approximation, the T matrix reads:

\begin{equation}
    T_{\vb{k}^{\prime}\sigma^{\prime},\vb{k}\sigma}=V_{\sigma^{\prime},\sigma}(\vb{k}^{\prime}-\vb{k})+\sum_{\vb{k}^{\prime\prime}\sigma^{\prime\prime}}
    \frac{V_{\sigma^{\prime},\sigma^{\prime\prime}}(\vb{k}^{\prime}-\vb{k}^{\prime\prime})V_{\sigma^{\prime\prime},\sigma}(\vb{k}^{\prime\prime}-\vb{k})}{\epsilon_{\vb{k},\sigma}-\epsilon_{\vb{k}^{\prime\prime},\sigma^{\prime\prime}}+i\eta}
    \label{eq:3}
\end{equation}
where $V_{\sigma^{\prime},\sigma}(\vb{k}^{\prime}-\vb{k})=\mel{\vb{k}^{\prime}\sigma^{\prime}}{V}{\vb{k}\sigma}$ is the transition matrix element from the initial state$\ket*{\vb{k}\sigma}$ to the final state$\ket*{\vb{k}^{\prime}\sigma^{\prime}}$, $\eta$ is a positive infinitesimal that specifies the retarded boundary condition, and $\epsilon_{\vb{k}\sigma/\vb{k}^{\prime}\sigma^{\prime}} $is the energy of the initial or final state. The scattering probability is given by:

\begin{widetext}
\begin{align}
P_{\vb{k}^{\prime}\sigma^{\prime},\vb{k}\sigma}
&=
2\pi n_{\mathrm{imp}}
\left|
T_{\vb{k}^{\prime}\sigma^{\prime},\vb{k}\sigma}
\right|^2
\delta\!\left(
\epsilon_{\vb{k}}-\epsilon_{\vb{k}^{\prime}}
\right)
\nonumber
\\
&=
2\pi n_{\mathrm{imp}}
\delta\!\left(
\epsilon_{\vb{k}}-\epsilon_{\vb{k}^{\prime}}
\right)
\left[
\left|
V_{\sigma,\sigma^{\prime}}
\left(\vb{k}-\vb{k}^{\prime}\right)
\right|^2
+
\sum_{\vb{k}^{\prime\prime},\sigma^{\prime\prime}}
\frac{
V_{\sigma^{\prime},\sigma^{\prime\prime}}
\left(\vb{k}^{\prime}-\vb{k}^{\prime\prime}\right)
V_{\sigma^{\prime\prime},\sigma}
\left(\vb{k}^{\prime\prime}-\vb{k}\right)
V_{\sigma,\sigma^{\prime}}
\left(\vb{k}-\vb{k}^{\prime}\right)
}{
\epsilon_{\vb{k},\sigma}
-
\epsilon_{\vb{k}^{\prime\prime},\sigma^{\prime\prime}}
+i\eta
}
\right].
\label{eq:4}
\end{align}
\end{widetext}
where $n_{imp}$ is the impurity density and we set $\hbar=1$.\par

In any order, the transition probability $P_{\vb{k}^{\prime}\sigma^{\prime},\vb{k}\sigma}$ can be decomposed to a symmetric part and an anti-symmetric part. Based on Eq.\ref{eq:4}, the anti-symmetric part emerges when considering the $V^3$ order dependence of the transition probability. Owing to spin-orbit coupling (SOC), the transition probability becomes spin-dependent. Consequently, the scattering distributions differ for spin-up and spin-down particles, leading to the emergence of a net spin current. \par

Although the argument presented above is based on Born's approximation, it accurately represents the main result of skew scattering. Generally, the spin dependent scattering probability can be divided into two parts as follows:
\begin{equation}
    P_{\vb{k}^{\prime},\vb{k}}=I_{\vb{k}^{\prime},\vb{k}}\sigma_{0}+I_{\vb{k}^{\prime},\vb{k}}S_{\vb{k}^{\prime},\vb{k}}\vb{\sigma}\cdot \vb{n}
    \label{eq:5}
\end{equation}
where $\vb{n}=\hat{\vb{k}}\times \hat{\vb{k}^{\prime}}$ is the unit vector perpendicular to the scattering plane and $S$ is the Sherman function\cite{PhysRevLett.95.166605}, which measures the polarization of outgoing electrons scattered into the direction $\vb{k}^{\prime}$ from an unpolarized incoming beam of momentum $\vb{k}$. And $I_{\vb{k}^{\prime},\vb{k}}$ measures the spin-independent part of scattering amplitude. \par

In 2D systems, both intrinsic and extrinsic SOC contribute to the spin Hall effect. However, the minimal model that encompasses the skew scattering contribution solely involves intrinsic SOC. Ferreira and Pappoport \cite{PhysRevLett.112.066601} suggest that the general structure of the T-Matrix in 2D systems with time-reversal and $C_2$ symmetries is given by:
\begin{equation}
    T_{\vb{k}^{\prime},\vb{k}}=a\sigma_{0}+(b\sigma_z+c \vec{n}\cdot \vb{\sigma})\sin{\theta}
    \label{eq:6}
\end{equation}
where $T_{\vb{k}^{\prime},\vb{k}}$is a $2\times2$ matrix with spin indices, $\theta$ is the angle between $\vb{k}$ and $\vb{k}^{\prime}$, and $\vb{n}=\hat{\vb{k}}\times \hat{\vb{k}^{\prime}}$. a represents the strength of spin-independent scattering, b represents skew scattering, and c represents spin-flip scattering.\par
The symmetric part and the spin-dependent skew part of the electron distribution function have different effective relaxation time. Previous work has shown that for the first-order skew part of the electron distribution function, the effective relaxation time is :
\begin{equation}
    \frac{1}{\tau_{sk}}=\int d\Omega I(\theta)S(\theta)\sin{\theta}
    \label{eq:7}
\end{equation}
\cite{PhysRevLett.95.166605},which measures the strength of skew scattering. We will calculate the higher-order corrections to the distribution function in a more general way in section $\mathrm{\ref{sec:Method}}$.

\section{\label{sec:Method} BOLTZMANN TRANSPORT EQUATION}
\subsection{\label{sec:m1} Calculation of distribution function }
Present theories of nonlinear charge and spin current are mainly based on the relaxation time approximation. However, in our case with skew scattering, the single-relaxation-time approximation is generally insufficient because the symmetric and skew scattering channels act differently on the angular harmonics of the distribution. We therefore retain the full collision integral of the Boltzmann transport equation (BTE):
\begin{equation}
    \grad_{\vb{k}}f_{\sigma}(\vb{k})\cdot (-e \vb{E})=\sum_{\vb{k'},\sigma'}(f_{\sigma'}(\vb{k'})P_{\vb{k}\sigma,\vb{k'}\sigma'}-f_\sigma(\vb{k})P_{\vb{k'}\sigma',\vb{k}\sigma})
    \label{eq:8}
\end{equation}
Our work focuses on the properties of nonlinear spin Hall transport in two-dimensional systems. Following previous work, we assume that the transition probability $P_{\vb{k'},\vb{k}}$ depends solely on the angle between the momenta before and after the collision: $P_{\vb{k'},\vb{k}}=P(\theta'-\theta)$, where $\theta,\theta'$ is the angle of $\vb{k},\vb{k'}$ in polar coordinate system. Since electronic transitions are restricted to the same energy surface, the electron distribution function becomes solely dependent on the angle for a given energy. Therefore, the collision terms in Boltzmann equation are composed of the following integrations : 
\begin{equation}
    \int_{0}^{2\pi} \frac{d\theta'}{2\pi} I(\theta-\theta')f(\theta')
    \label{eq:11}
\end{equation}
The Boltzmann equation is an integral equation governing the electron distribution function, which can be converted into an algebraic equation through Fourier decomposition.
\begin{equation}
    f(\theta)=\sum_{l}f_l e^{i l\theta}
    \label{eq:12}
\end{equation}
The integral in Eq.~\ref{eq:11} can be written as:
\begin{align}
 &=\sum_{l}f_l\int_{0}^{2\pi}\frac{d\theta'}{2\pi}I(\theta-\theta')e^{i l \theta'}\nonumber\\
    &=\sum_{m}f_le^{i l \theta}\int_{0}^{2\pi}\frac{d\theta'}{2\pi}I(\theta-\theta')e^{-i l(\theta-\theta')}\nonumber\\
    &=\sum_l f_lI_l e^{il\theta}
\end{align}
,where $I_l=\int_0^{2\pi}\frac{d\alpha}{2\pi}I(\alpha)e^{-i l \alpha}$ is the $l$th Fourier expansion coefficient of $I(\theta)$.\\
Based on the set of algebraic equations satisfied by the Fourier components of the electron distribution function, we can calculate its response to an external electric field. It is crucial to note that each Fourier component is characterized by a distinct relaxation time, rather than being described by a single, unified timescale. We use the surface states of topological insulators as a platform to demonstrate our techniques.\\

The surface Hamiltonian is:
\begin{equation}
    H_{\mathrm{TI}}=\frac{k^2}{2m}+v(k_x\sigma_y-k_y\sigma_x)+\Delta\sigma_z
    \label{eq:9}
\end{equation}
where v is the velocity for the Dirac cone and $\Delta$ is the possible gap opened by the Zeeman field, interactions or other mechanisms. The energy dispersion is $\epsilon^{\pm}=\frac{k^2}{2m}\pm h$,where $h=\sqrt{v^2k^2+\Delta^2}$ and $\pm$ represent the upper and lower bands. The normalized eigenstates of the Hamiltonian are:

\begin{equation}
    u_{\pm}(k,\theta)=\frac{1}{\sqrt{k^2v^2+(\Delta\mp h)^2}}\mqty(i vk e^{-i\theta}\\  \Delta\mp h)
    \label{eq:10}
\end{equation}
The T-matrix of the system takes the form of equation \ref{eq:6} and we assume that $c=0$, a and b are constant. Keeping only the leading contribution in the skew-scattering amplitude b, the transition probability between state $\ket{u_{\alpha}} $ and $\ket{u_{\beta}}$ is :
\begin{equation}
    \abs{T_{\beta\alpha}}^2=a^2\abs{u_{\alpha}^{\dag}u_{\beta}}^2+2ab\mathrm{Re}[u_{\alpha}^{\dag}u_{\beta}u_{\beta}^{\dag}\sigma_{z}u_{\alpha}]\sin\theta+O(b^2)
    \label{eq:13}
\end{equation}
, where $\theta=\theta_2-\theta_1$ is the difference between the angles of the initial and final state.
According to the eigenstates in Eq.\ref{eq:10}, the transition probability can be divided into two parts:
$I_{\beta \alpha}(\theta)\equiv a^2\abs{u_{\alpha}^{\dag}u_{\beta}}^2$,which is an even function of $\theta$, represents the symmetric scattering part and $G_{\beta \alpha}(\theta)\equiv2ab\mathrm{Re}[u_{\alpha}^{\dag}u_{\beta}u_{\beta}^{\dag}\sigma_{z}u_{\alpha}]\sin\theta$,which is an odd function of $\theta$, represents the skew scattering part.\\

The system is considered in a configuration where  uniform electric field $\vb{E}=E\vb{e_x}$ is applied along the x-axis. The electron distribution function can be expanded by the order of the electric field E:
\begin{equation}
    f=f_{sym}^{(0)}+f_{sym}^{(1)}+f_{sk}^{(1)}+f_{sym}^{(2)}+f_{sk}^{(2)}+\cdots
\end{equation}
, where $f_{sym}^{(0)}=\frac{1}{e^{\beta(\epsilon-\mu)}+1}$ is the electron distribution without an electric field and $(n)$  represents the perturbative order and the power-law dependence of the distribution function on the electric field E.\\
Therefore, the Boltzmann equation can be organized by the order of the electric field, the symmetric part of the Boltzmann equation can be written as:
\begin{widetext}
    \begin{align}
    -e\vb{E}\cdot\pdv{f_{+,sym}^{(n)}(k_{+},\theta)}{\vb{k_{+}}}&=\iint \frac{k'dk'd\theta'}{(2\pi)^2}2\pi\delta(\epsilon_k-\epsilon_{k'})[I_{+-}(\theta-\theta')f_{-,sym}^{(n+1)}(k',\theta')-I_{-+}(\theta-\theta')f_{+,sym}^{(n+1)}(k_{+},\theta)\nonumber\\ &+I_{++}(\theta-\theta')f_{+,sym}^{(n+1)}(k',\theta')-I_{++}(\theta-\theta')f_{+,sym}^{(n+1)}(k_{+},\theta)]\nonumber \\
    &=\int_{0}^{2\pi}\frac{d\theta'}{2\pi}(\frac{m}{1-\frac{mv^2}{h_{-}}}(I_{+-}(\theta-\theta')f_{-,sym}^{(n+1)}(k_{-},\theta')-I_{-+}(\theta-\theta')f_{+,sym}^{(n+1)}(k_{+},\theta))\nonumber\\&+\frac{m}{1+\frac{mv^2}{h_{+}}}(I_{++}(\theta-\theta')f_{+,sym}^{(n+1)}(k_{+},\theta')-I_{++}(\theta-\theta')f_{+,sym}^{(n+1)}(k_{+},\theta))
    \label{eq:14}
    \end{align}
    
    \begin{align}
        -e\vb{E}\cdot\pdv{f_{-,sym}^{(n)}(k_{-},\theta)}{\vb{k_{-}}}&=\iint \frac{k'dk'd\theta'}{(2\pi)^2}2\pi\delta(\epsilon_k-\epsilon_{k'})[I_{-+}(\theta-\theta')f_{+,sym}^{(n+1)}(k',\theta')-I_{+-}(\theta-\theta')f_{-,sym}^{(n+1)}(k_{-},\theta)\nonumber\\ &+I_{--}(\theta-\theta')f_{-,sym}^{(n+1)}(k',\theta')-I_{--}(\theta-\theta')f_{-,sym}^{(n+1)}(k_{-},\theta)]\nonumber \\
    &=\int_{0}^{2\pi}\frac{d\theta'}{2\pi}(\frac{m}{1+\frac{mv^2}{h_{+}}}(I_{-+}(\theta-\theta')f_{+,sym}^{(n+1)}(k_{+},\theta')-I_{+-}(\theta-\theta')f_{-,sym}^{(n+1)}(k_{-},\theta)\nonumber\\&+\frac{m}{1-\frac{mv^2}{h_{-}}}(I_{--}(\theta-\theta')f_{-,sym}^{(n+1)}(k_{-},\theta')-I_{--}(\theta-\theta')f_{-,sym}^{(n+1)}(k_{-},\theta))
    \label{eq:15}
    \end{align}
\end{widetext}
,where $h_{\pm}=\sqrt{k_{\pm}^2v^2+\Delta^2}$ and $k_{+},k_{-}$ satisfy $\epsilon_{+}(k_{+})=\epsilon_{-}(k_{-})$.
We work to leading order in the antisymmetric skew-scattering rate and neglect the contribution obtained by applying the skew-scattering part of the collision operator to $f_{\mathrm{sk}}$. In the weak-skew-scattering regime, $f_{\mathrm{sk}}$ is itself first order in the antisymmetric scattering rate, so this contribution is second order in the skew-scattering strength and lies beyond the accuracy retained here.The Fourier expansion of the equation can be written in matrix form:
\begin{widetext}
\begin{equation}
    \mqty[\frac{m}{1+\frac{mv^2}{h_{+}}}(I_{++,l}-I_{++,0})-\frac{m}{1-\frac{mv^2}{h_{-}}}I_{+-,0}& \frac{m}{1-\frac{mv^2}{h_{-}}}I_{+-,l}\\
    \frac{m}{1+\frac{mv^2}{h_{+}}}I_{-+,l}&\frac{m}{1-\frac{mv^2}{h_{-}}}(I_{--,l}-I_{--,0})-\frac{m}{1+\frac{mv^2}{h_{+}}}I_{-+,0}]
    \mqty[f_{+,sym,l}^{(n+1)}(k_{+})\\f_{-,sym,l}^{(n+1)}(k_{-})]=\mqty[[-e\vb{E}\cdot\pdv{f_{+,sym}^{(n)}(k_{+},\theta)}{\vb{k_{+}}}]_{l}\\
    [-e\vb{E}\cdot\pdv{f_{-,sym}^{(n)}(k_{-},\theta)}{\vb{k_{-}}}]_{l}
    ]
    \label{eq:16}
\end{equation}
\end{widetext}
The skewing part of the Boltzmann equation has a similar structure to the symmetric part, except that the collision term contains two types of contributions: the first is $f_{sk}I(\theta)$, the second is $f_{sym}G(\theta)$. Therefore, the Fourier expansion of the skew-scattering part of the Boltzmann Equation can be written as:
\begin{widetext}
\begin{align}
\mqty[[-e\vb{E}\cdot\pdv{f_{+,sk}^{(n)}(k_{+},\theta)}{\vb{k_{+}}}]_{l}\\
    [-e\vb{E}\cdot\pdv{f_{-,sk}^{(n)}(k_{-},\theta)}{\vb{k_{-}}}]_{l}
    ]&=
        \mqty[\frac{m}{1+\frac{mv^2}{h_{+}}}(I_{++,l}-I_{++,0})-\frac{m}{1-\frac{mv^2}{h_{-}}}I_{+-,0}& \frac{m}{1-\frac{mv^2}{h_{-}}}I_{+-,l}\\
    \frac{m}{1+\frac{mv^2}{h_{+}}}I_{-+,l}&\frac{m}{1-\frac{mv^2}{h_{-}}}(I_{--,l}-I_{--,0})-\frac{m}{1+\frac{mv^2}{h_{+}}}I_{-+,0}]
    \mqty[f_{+,sk,l}^{(n+1)}(k_{+})\\f_{-,sk,l}^{(n+1)}(k_{-})]\nonumber\\
    &+\mqty[\frac{m}{1+\frac{mv^2}{h_{+}}}G_{++,l}& \frac{m}{1-\frac{mv^2}{h_{-}}}G_{+-,l}\\
    \frac{m}{1+\frac{mv^2}{h_{+}}}G_{-+,l}&\frac{m}{1-\frac{mv^2}{h_{-}}}G_{--,l}]\mqty[f_{+,sym,l}^{(n+1)}(k_{+})\\f_{-,sym,l}^{(n+1)}(k_{-})]
    \label{eq:17}
\end{align}
We have utilized that $G(\theta)$ is an odd function of $\theta$, so there is no zeroth Fourier component.
\end{widetext}
Based on Eqs.~\eqref{eq:16} and \eqref{eq:17}, the nonequilibrium distribution function can be determined recursively order by order in the applied electric field. Starting from the equilibrium Fermi--Dirac distribution, the drift term generated by the $n$th-order correction provides the source term for the $(n+1)$th-order Boltzmann equation. Decomposing this equation into angular Fourier harmonics reduces the collision integral to a set of band-coupled algebraic equations, from which the symmetric and skew-scattering components of the $(n+1)$th-order distribution can be obtained for each harmonic. In the following, we apply this recursive procedure up to second order in the electric field.
\begin{figure*}[t] \centering \includegraphics[width=\textwidth]{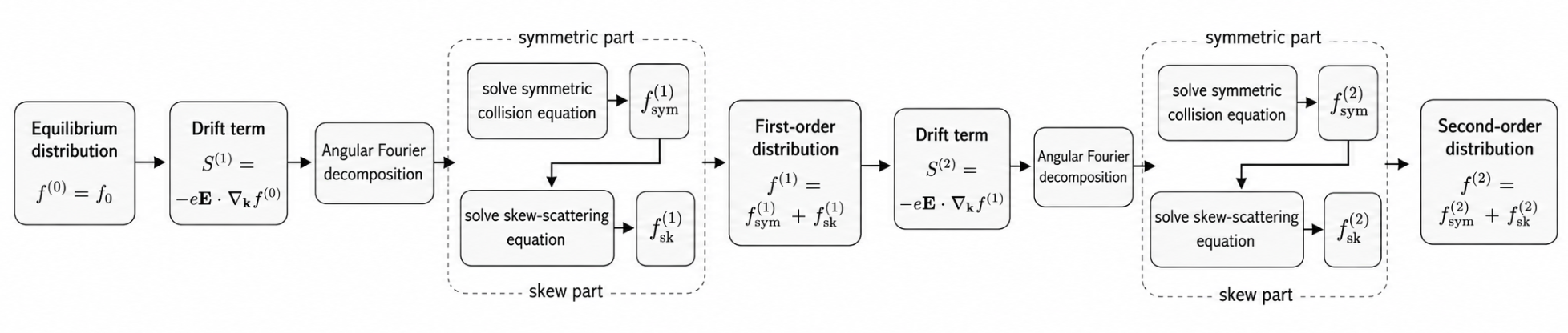} \caption{ Recursive procedure for solving the Boltzmann equation order by order in the applied electric field.} \label{fig:recursive_boltzmann} \end{figure*}
The recursive procedure for obtaining the nonequilibrium distribution function is summarized in Fig.~\ref{fig:recursive_boltzmann}.
\subsection{Spin Current}
Our primary interest is the extrinsic contribution to the nonlinear spin current arising from the nonequilibrium distribution function obtained from the full collision integral. Within the relaxation-time approximation, the second-order extrinsic spin-current response coefficient is given by
\begin{equation}
    \Gamma^{\ell,ext}_{ijl}=e^2\int_{\vb{k}}[j_{i}^{\ell}\frac{\tau^2}{\hbar^2}\pdv{f^{(0)}}{k_j}{k_l}-\frac{\tau}{\hbar}\Omega_{ij}^{\ell}\pdv{f^{(0)}}{k_{l}}]
\end{equation}
,where
\begin{equation}
    j_{i}^{\ell}=\mel{u_{n}^{(0)}}{\hat{j_{i}^{\ell}}}{u_{n}^{(0)}}=\mel{u_{n}^{(0)}}{\frac{1}{4}\{\pdv{H}{k_i},\sigma^{\ell}\}}{u_{n}^{(0)}}
\end{equation}
\begin{equation}
    \Omega_{ij}^{\ell}=\sum_{m\neq n}2\mathrm{Im}\left[\frac{\mel{u_{n}^{(0)}}{\frac{1}{4}\{\pdv{H}{k_i},\sigma^{\ell}\}}{u_{m}^{(0)}}\mel{u_{m}^{(0)}}{\pdv{H}{k_j}}{u_{n}^{(0)}}}{(\epsilon_{n}^{(0)}-\epsilon_{m}^{(0)})^2}\right]
    \label{eq:18}
\end{equation}
\cite{PhysRevB.110.174434}
$j_i^\ell$ is the conventional spin-current matrix element and $\Omega_{ij}^{\ell}$ is the corresponding spin Berry curvature. In the present formulation, the relaxation-time expressions
$
\frac{e\tau}{\hbar}\frac{\partial f^{(0)}}{\partial k_l}
\quad\text{and}\quad
\frac{e^2\tau^2}{\hbar^2}
\frac{\partial^2 f^{(0)}}{\partial k_j\partial k_l}
$
are replaced by the explicitly calculated first- and second-order distribution functions, $f_l^{(1)}$ and $f_{jl}^{(2)}$, respectively. 
\begin{equation}
    \Gamma_{ijl}^{\ell,ext}=\int_{\vb{k}}(j_{i}^{\ell}f_{jl}^{(2)}-e\Omega^{\ell}_{ij}f_{l}^{(1)})
\end{equation}
These distribution functions are obtained from the recursive solution of the Boltzmann equation and retain the angular and band dependence of the collision integral.

\section{\label{sec:result}Main Results}
We calculated the extrinsic part of the second-order conductivities for spin current in the model Hamiltonian: $H_{TI}=\frac{k^2}{2m}+v(k_x\sigma_y-k_y\sigma_x)+\Delta\sigma_z$
with T-matrix:$T_{\vb{k},\vb{k'}}=a \sigma_0+b\sigma_z\sin\theta$. 
For the numerical calculations, all physical quantities are expressed in dimensionless form using the elementary charge $e$, the reduced Planck constant $\hbar$, the characteristic energy scale $E_0=0.1\mathrm{eV}$, and the characteristic momentum scale $k_0=0.01\,\mathrm{\AA}^{-1}$. In particular, we define
$
\tilde{\mathbf{k}}=\frac{\mathbf{k}}{k_0},
\quad
\tilde{\varepsilon}=\frac{\varepsilon}{E_0},
\quad
\tilde{\mu}=\frac{\mu}{E_0},
$
and set $e=\hbar=E_0=k_0=1$ in the numerical evaluation. The corresponding electric-field scale is
$
\mathcal{E}_0=\frac{E_0k_0}{e}.
$
For the two-dimensional spin-current density used here, the natural scale is $J_0=E_0k_0$. We therefore report the second-order spin-current response coefficient in units of
$
\Gamma_0=\frac{J_0}{\mathcal{E}_0^2}
=\frac{e^2}{E_0k_0}.
$
Unless otherwise stated, all quantities shown in the figures are dimensionless.\par
The qualitative structure of the lower band is controlled by the ratio $mv^2/\Delta$. For $mv^2/\Delta>1$, the lower band is nonmonotonic and develops a minimum at finite momentum. Consequently, a finite energy interval contains two constant-energy contours in the lower band. By contrast, for $mv^2/\Delta<1$, the lower band increases monotonically with $k$, and only one lower-band contour is present at a given energy. The two representative parameter sets considered below are
\begin{align}
&(a,b,m,v,\Delta,\beta)=(1,0.2,2,5,2,5),\\
&(a,b,m,v,\Delta,\beta)=(1,0.2,2,1,5,5),
\end{align}
which lie in the nonmonotonic and monotonic regimes, respectively.

\begin{figure}[t]
    \centering

    \begin{overpic}[width=0.48\linewidth,height=0.14\textheight]{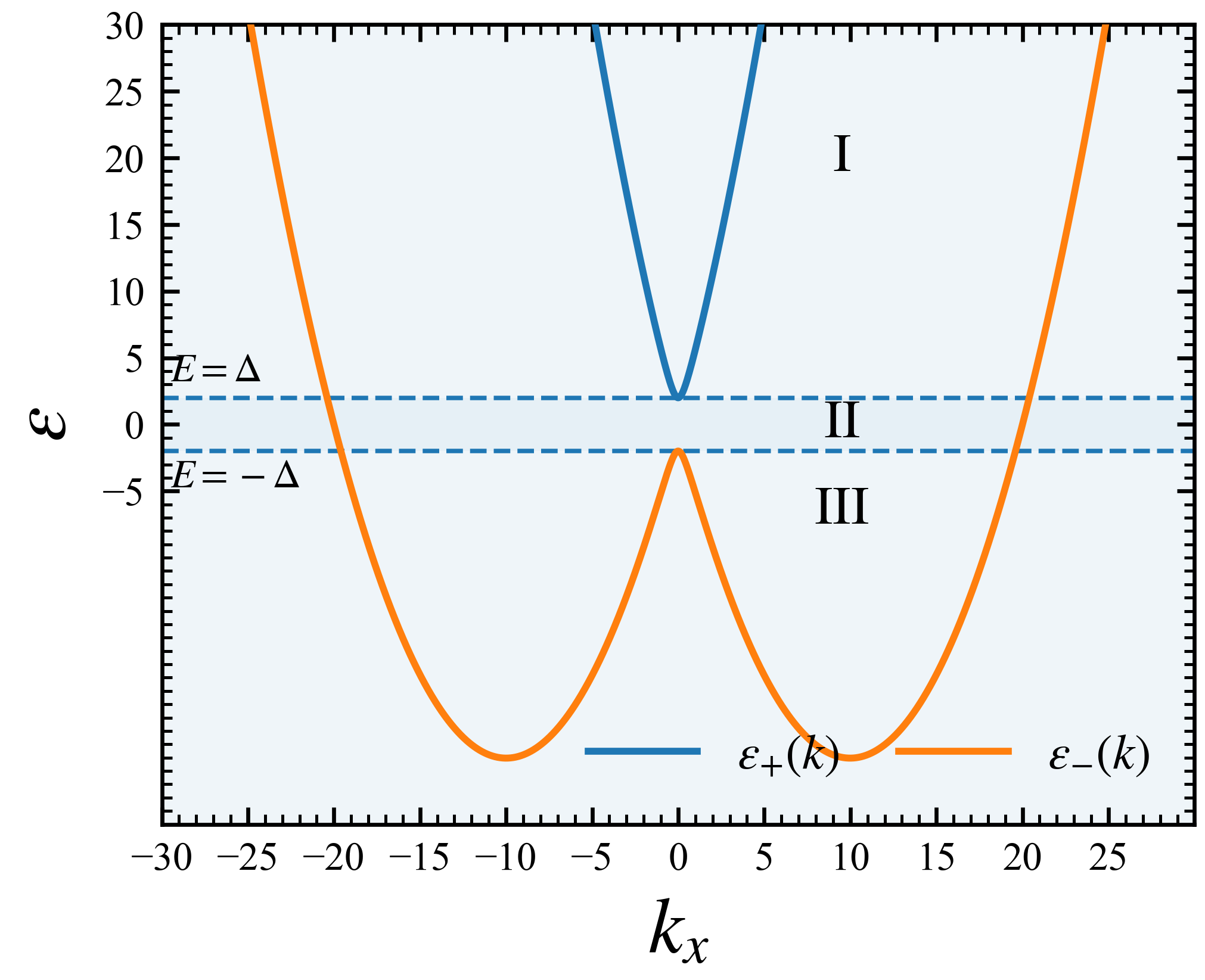}
        \put(-2,90){\small (a)}
    \end{overpic}%
    \hfill
    \begin{overpic}[width=0.48\linewidth,height=0.14\textheight]{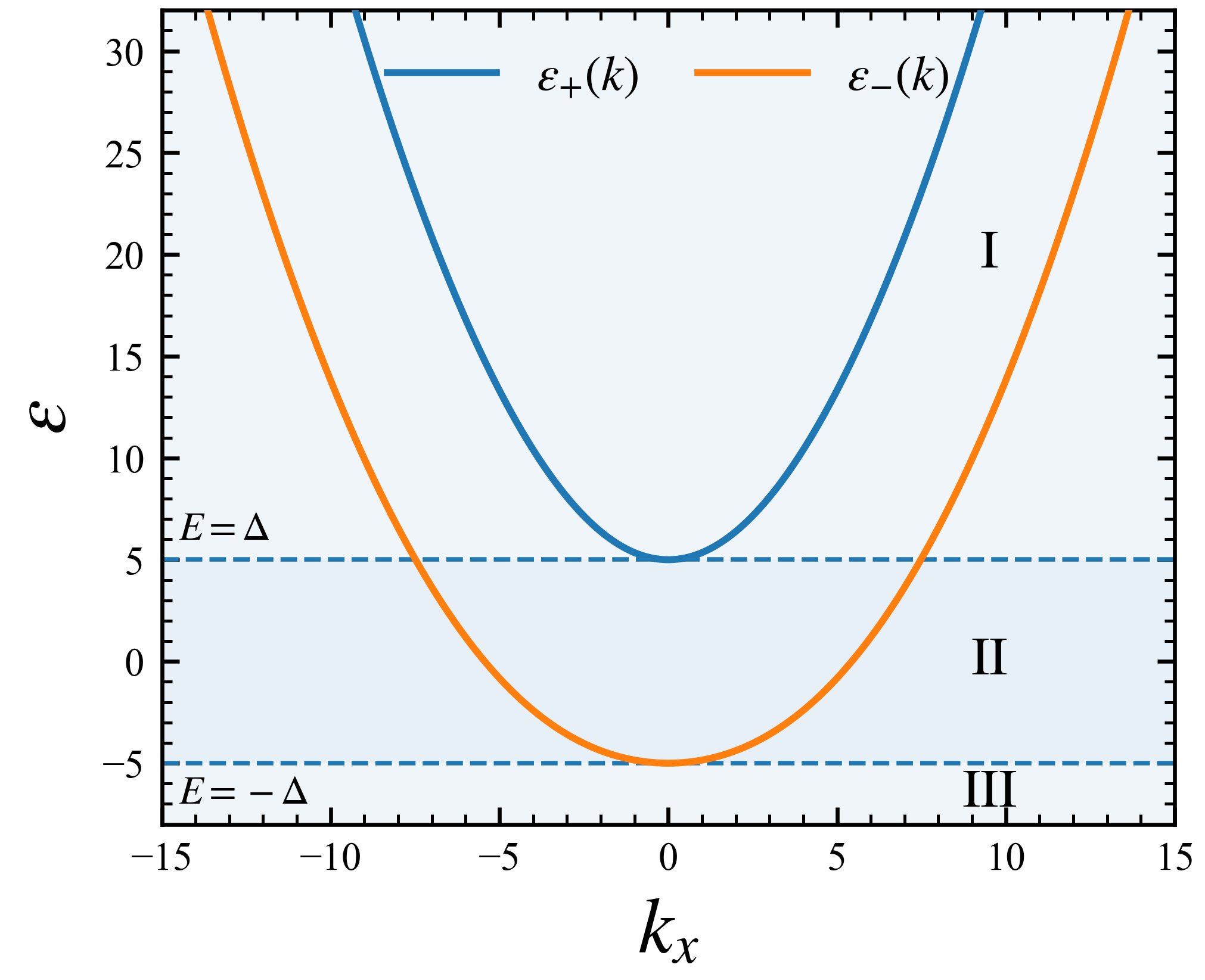}
        \put(-2,90){\small (b)}
    \end{overpic}

    \caption{
    Energy-band structures of the Hamiltonian describing the surface states
    of a topological insulator for
    (a)~$v=5$, $\Delta=2$, and $m=2$;
    (b)~$v=1$, $\Delta=5$, and $m=2$.
    }
    \label{fig:band}
\end{figure}

Fig.~\ref{fig:band} shows the corresponding band structures. In the nonmonotonic regime, three energy regions can be distinguished according to the available elastic-scattering channels. In region I, states in the upper and lower bands coexist at the same energy, so both intraband and interband scattering processes are allowed. In region II, only one constant-energy contour of the lower band is present. In region III, the lower band supports inner and outer constant-energy contours, allowing scattering both within each contour and between the two contours. Although the explicit collision matrices differ among these regions, they are treated within the same angular-harmonic framework developed in Sec.~III.

As shown in Appendix \ref{sec:appendixB},the spin-current matrix elements contain only zeroth and second angular harmonics. Owing to the orthogonality of the angular Fourier modes, the angular integration selects the corresponding components of the nonequilibrium distribution function. In particular, the anisotropic part of the second-order spin-current response is determined by the $l=2$ harmonic of the distribution. This harmonic selection clarifies why the angular dependence of the collision operator is important for nonlinear spin transport. The conventional transport relaxation time is primarily defined by the relaxation of the $l=1$ angular mode, which controls the linear current. At second order in the electric field, the drift term generates $l=0$ and $l=2$ components of the distribution. Although a constant-$\tau$ approximation can formally produce these higher harmonics, it assigns them the same phenomenological relaxation time as the $l=1$ mode. In the full collision-integral treatment, by contrast, each angular harmonic is governed by its own band-resolved collision matrix. The $l=2$ component relevant to the nonlinear spin current may therefore relax differently from the linear $l=1$ mode and may additionally be coupled through interband or inter-contour scattering. These effects are not captured by a single-relaxation-time approximation.

\begin{figure}[t]
    \centering

    \begin{overpic}[width=0.48\linewidth,height=0.14\textheight]{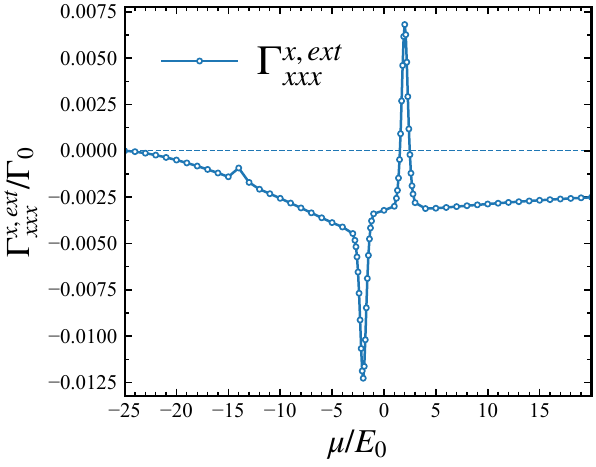}
        \put(2,80){\small (a)}
    \end{overpic}%
    \hfill
    \begin{overpic}[width=0.48\linewidth,height=0.14\textheight]{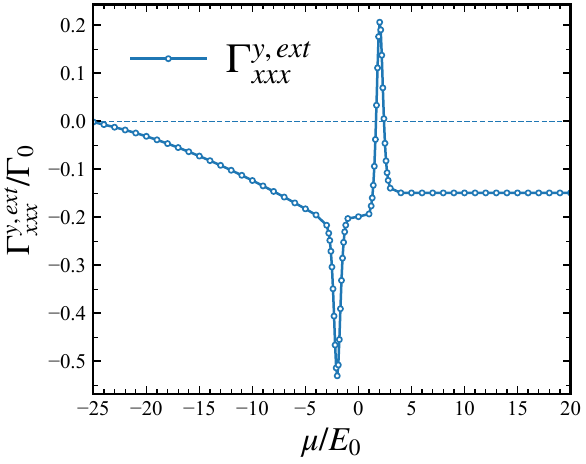}
        \put(2,80){\small (b)}
    \end{overpic}

    \vspace{0.6em}

    \begin{overpic}[width=0.48\linewidth,height=0.14\textheight]{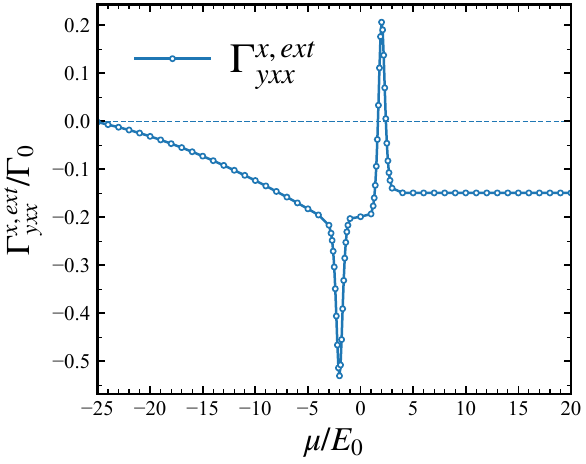}
        \put(2,80){\small (c)}
    \end{overpic}%
    \hfill
    \begin{overpic}[width=0.48\linewidth,height=0.14\textheight]{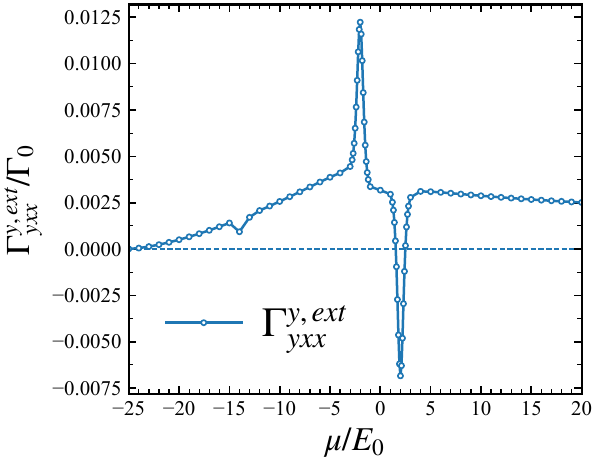}
        \put(2,80){\small (d)}
    \end{overpic}

    \caption{
    Second-order extrinsic nonlinear spin-current response coefficient
    $\Gamma_{knl}^{m,\mathrm{ext}}$ as a function of the chemical potential
    $\mu$ for $a=1$, $b=0.2$, $m=2$, $\Delta=2$, $v=5$, and $\beta=5$:
    (a)~$\Gamma_{xxx}^{x,\mathrm{ext}}$,
    (b)~$\Gamma_{xxx}^{y,\mathrm{ext}}$,
    (c)~$\Gamma_{yxx}^{x,\mathrm{ext}}$, and
    (d)~$\Gamma_{yxx}^{y,\mathrm{ext}}$.
    }
    \label{fig:gamma}
\end{figure}

Fig.~\ref{fig:gamma} presents the second-order extrinsic spin-current coefficients in the nonmonotonic regime. When the chemical potential lies below the minimum of the lower band, $\mu<\varepsilon_{\min}\simeq-25.0$, the occupation of the model bands is exponentially suppressed and all response coefficients approach zero. As the chemical potential enters the lower band, finite nonlinear spin currents develop. Pronounced structures appear near $\mu=-\Delta$ and $\mu=+\Delta$, where the topology and number of the available constant-energy contours change.

These sharp features should not be attributed solely to the second derivative of the Fermi distribution. The second-order distribution contains terms involving both derivatives of the equilibrium occupation and momentum derivatives of the collision matrices. Near the characteristic energies $\mu=\pm\Delta$, the density of states, group velocities, spin-current matrix elements, and available scattering channels vary rapidly. Their combined variation enhances the sensitivity of the momentum integral to states close to the Fermi energy and produces peaks of opposite sign in several tensor components. In particular, the feature near $\mu=-\Delta$ is associated with the change between the single-contour and two-contour sectors of the lower band, whereas the feature near $\mu=+\Delta$ coincides with the onset of upper-band states and additional interband scattering channels. For chemical potentials sufficiently above the upper-band edge, the response coefficients vary only weakly with $\mu$. 

\begin{figure}[t]
    \centering

    \begin{overpic}[width=0.48\linewidth,height=0.14\textheight]{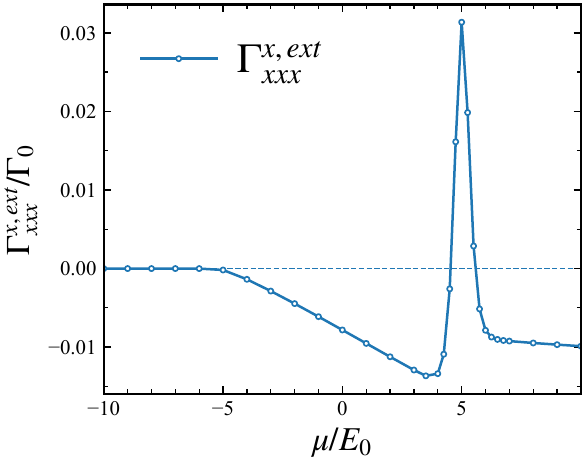}
        \put(2,80){\small (a)}
    \end{overpic}%
    \hfill
    \begin{overpic}[width=0.48\linewidth,height=0.14\textheight]{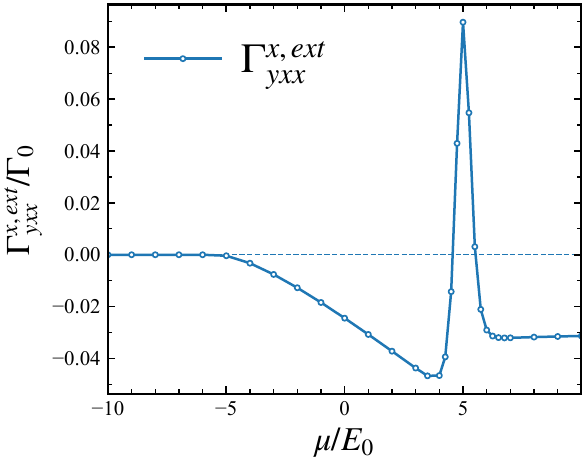}
        \put(2,80){\small (b)}
    \end{overpic}

    \vspace{0.6em}

    \begin{overpic}[width=0.48\linewidth,height=0.14\textheight]{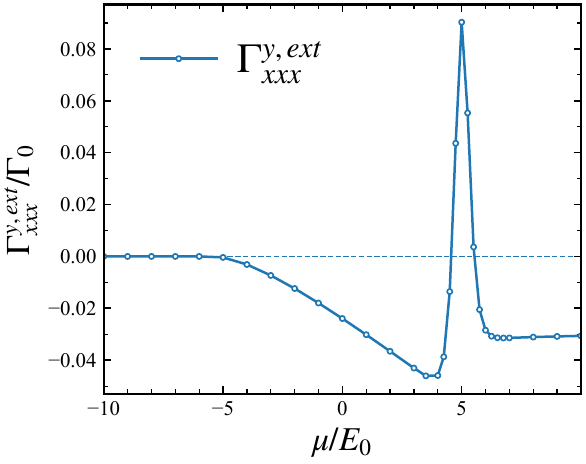}
        \put(2,80){\small (c)}
    \end{overpic}%
    \hfill
    \begin{overpic}[width=0.48\linewidth,height=0.14\textheight]{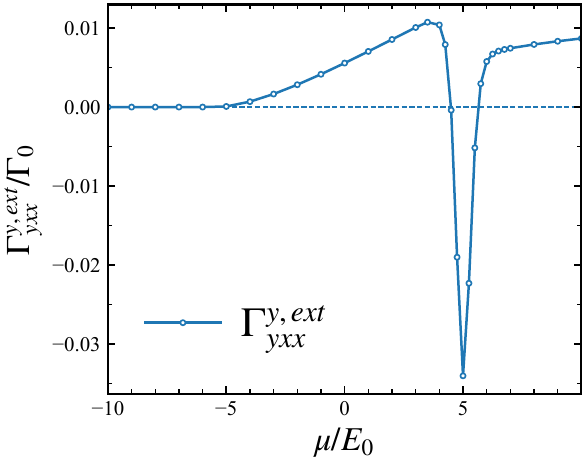}
        \put(2,80){\small (d)}
    \end{overpic}

    \caption{
    Second-order extrinsic nonlinear spin-current response coefficient
    $\Gamma_{knl}^{m,\mathrm{ext}}$ as a function of the chemical potential
    $\mu$ for $a=1$, $b=0.2$, $m=2$, $\Delta=5$, $v=1$, and $\beta=5$:
    (a)~$\Gamma_{xxx}^{x,\mathrm{ext}}$,
    (b)~$\Gamma_{xxx}^{y,\mathrm{ext}}$,
    (c)~$\Gamma_{yxx}^{x,\mathrm{ext}}$, and
    (d)~$\Gamma_{yxx}^{y,\mathrm{ext}}$.
    }
    \label{fig:gamma2}
\end{figure}

The corresponding results for the monotonic regime are shown in Fig.~\ref{fig:gamma2}. Because the lower band contains only one constant-energy contour, the inner--outer-contour scattering processes present in the nonmonotonic regime are absent. The response consequently exhibits a simpler chemical-potential dependence. In particular, the pronounced structure near $\mu=-\Delta$ is removed, while a dominant peak remains near $\mu=+\Delta$, where the upper band becomes accessible and the set of elastic-scattering channels changes.

The comparison between the two parameter regimes demonstrates that the nonlinear spin-current response is controlled not only by the local spin texture of the eigenstates but also by the global structure of the constant-energy contours entering the collision integral. The qualitative change across $mv^2/\Delta=1$ therefore originates from the transition between a nonmonotonic lower band with two possible contours and a monotonic lower band with a single contour. This result illustrates the importance of treating the angular harmonics and contour-resolved scattering processes explicitly in extrinsic nonlinear spin transport.

The numerical results exhibit the approximate relations
\begin{equation}
    \Gamma^{x}_{xxx} \simeq -\Gamma^{y}_{yxx},
    \qquad
    \Gamma^{x}_{yxx} \simeq \Gamma^{y}_{xxx}.
\end{equation}
These relations can be understood from the angular structure of the
spin-current operators and the nonequilibrium distribution functions.
For the rotationally symmetric massive Dirac model considered here,
the spin-momentum locking leads to
\begin{equation}
    j_x^x = - j_y^y
\end{equation}
for the $l=2$ angular harmonics relevant to the transverse
spin-polarized response. More explicitly, both components are
proportional to $\sin 2\theta$ with opposite signs. Since the
second-order skew-scattering correction to the distribution function
also contains a $\sin 2\theta$ harmonic, their conventional
contributions therefore satisfy
\begin{equation}
    \left(\Gamma^{x}_{xxx}\right)_{j f^{(2)}}
    =
    -\left(\Gamma^{y}_{yxx}\right)_{j f^{(2)}}.
\end{equation}
The equality is only approximate for the full response because
$\Gamma^{x}_{xxx}$ can additionally receive a contribution from the
field-induced spin-Berry-curvature term
$\Omega^{x}_{xx} f^{(1)}$, whereas the corresponding
$\Omega^{y}_{yx}$ contribution vanishes in the present model.

Similarly, the $l=2$ parts of $j_y^x$ and $j_x^y$ are identical.
Indeed, writing only the angularly anisotropic components,
\begin{equation}
    \left(j_y^x\right)_{l=2}
    =
    \left(j_x^y\right)_{l=2}
    \propto \cos 2\theta .
\end{equation}
They therefore couple identically to the $\cos 2\theta$ component of
the second-order symmetric distribution, giving
\begin{equation}
    \left(\Gamma^{x}_{yxx}\right)_{j f^{(2)}}
    =
    \left(\Gamma^{y}_{xxx}\right)_{j f^{(2)}}.
\end{equation}
The small deviation between the two full response coefficients
originates from the spin-Berry-curvature contribution:
$\Omega^{x}_{yx}$ is finite and couples to the first-order
skew-scattering distribution, whereas
$\Omega^{y}_{xx}=0$. Consequently, the difference
$\Gamma^{x}_{yxx}-\Gamma^{y}_{xxx}$ is controlled by an additional
skew-scattering-induced correction.

For the parameters used here, $a=1$ and $b=0.2$, the symmetric
scattering channel remains dominant, while the finite $b$ produces a
moderate skew-scattering correction. The latter, together with the
spin-Berry-curvature contribution, weakly breaks the above relations,
explaining why the corresponding response coefficients are not
exactly equal (or opposite) but remain numerically very close over a
broad range of chemical potential. These approximate relations are
therefore not independent symmetry constraints on the nonlinear
response tensor, but rather reflect the combination of in-plane
rotational symmetry, spin-momentum locking, and angular-harmonic
selection in the present model.

\par

\section{\label{sec:scale} DISCUSSION OF DISORDER SCALING}

The disorder dependence of the nonlinear spin-current response can be inferred directly from the structure of the collision matrices. We denote the overall strengths of the symmetric and antisymmetric parts of the collision operator by $\rho_{\mathrm{sym}}$ and $\rho_{\mathrm{sk}}$, respectively, such that
\begin{equation}
M_l\propto \rho_{\mathrm{sym}}\text{,}
\quad
P_l\propto \rho_{\mathrm{sk}}\text{.}
\end{equation}
,where $M_l$ and $P_l$ denote the scattering matrices associated with the symmetric and skew-scattering parts of the transition probability, respectively, as defined in Appendix~A.
Here, $\rho_{\mathrm{sym}}$ characterizes ordinary momentum-relaxing scattering, whereas $\rho_{\mathrm{sk}}$ parametrizes the antisymmetric skew-scattering rate. The latter should be regarded as a scattering-strength scale rather than as an independently measurable longitudinal resistivity. The scaling then follows as:
\begin{equation}
f_{\mathrm{sym}}^{(1)}
\propto
\frac{1}{\rho_{\mathrm{sym}}}.
\end{equation}
The second-order symmetric component is generated by the drift term acting on $f_{\mathrm{sym}}^{(1)}$ and therefore obeys
\begin{equation}
f_{\mathrm{sym}}^{(2)}
\propto
M_2^{-1}f_{\mathrm{sym}}^{(1)}
\propto
\frac{1}{\rho_{\mathrm{sym}}^2}.
\end{equation}

To leading order in the antisymmetric scattering rate, the first-order skew correction is obtained from
\begin{equation}
M_1 f_{\mathrm{sk}}^{(1)}
\sim
P_1 f_{\mathrm{sym}}^{(1)},
\end{equation}
which gives
\begin{equation}
f_{\mathrm{sk}}^{(1)}
\propto
\frac{\rho_{\mathrm{sk}}}{\rho_{\mathrm{sym}}^2}.
\end{equation}
Similarly, the second-order skew correction receives contributions from the drift term acting on $f_{\mathrm{sk}}^{(1)}$ and from the antisymmetric collision operator acting on $f_{\mathrm{sym}}^{(2)}$. Both contributions have the same disorder scaling,
\begin{equation}
f_{\mathrm{sk}}^{(2)}
\propto
\frac{\rho_{\mathrm{sk}}}{\rho_{\mathrm{sym}}^3}.
\end{equation}
Therefore, the nonlinear spin-current obeys the scaling form: 
\begin{equation}
\frac{j_{a}^{b}}{E_{x}^2}=
\frac{A}{\rho_{\mathrm{sym}}^2}
+
\frac{B}{\rho_{\mathrm{sym}}}
+
C\frac{\rho_{\mathrm{sk}}}{\rho_{\mathrm{sym}}^3}
+
D\frac{\rho_{\mathrm{sk}}}{\rho_{\mathrm{sym}}^2}.
\label{eq}
\end{equation}
The coefficients $A$--$D$ depend on the band structure, chemical potential, temperature, and angular structure of the impurity potential, but are independent of the overall strengths $\rho_{\mathrm{sym}}$ and $\rho_{\mathrm{sk}}$.
In the weak-skew-scattering regime considered here, the symmetric scattering contribution dominates the longitudinal momentum relaxation,
\begin{equation}
\rho_{\mathrm{sk}}\ll \rho_{\mathrm{sym}}.
\end{equation}
We therefore neglect the correction of the antisymmetric scattering channel to the longitudinal charge conductivity. Within the Drude regime, the longitudinal electric field and charge current are approximately related by
\begin{equation}
E_x\simeq \rho_{\mathrm{sym}}j_x,
\end{equation}
or equivalently,
\begin{equation}
j_x\simeq \frac{E_x}{\rho_{\mathrm{sym}}}.
\end{equation}
It follows that
\begin{equation}
j_x^2\simeq\frac{E_x^2}{\rho_{\mathrm{sym}}^2}.
\end{equation}
Finally, we obtain:
\begin{equation}
\frac{j_{b}^{a}}{j_{x}^2}=   
A
+
B\rho_{\mathrm{sym}}
+
C\frac{\rho_{\mathrm{sk}}}{\rho_{\mathrm{sym}}}
+
D\rho_{\mathrm{sk}}.
\label{eq:scale1}
\end{equation}

Equation~\eqref{eq:scale1} provides a disorder-scaling relation for the nonlinear spin-current response normalized by the square of the longitudinal charge current. The four terms originate, respectively, from the contributions involving $f_{\mathrm{sym}}^{(2)}$, $f_{\mathrm{sym}}^{(1)}$, $f_{\mathrm{sk}}^{(2)}$, and $f_{\mathrm{sk}}^{(1)}$. In principle, their different dependences on $\rho_{\mathrm{sym}}$ and $\rho_{\mathrm{sk}}$ can be used to distinguish the symmetric and skew-scattering contributions by varying the concentration and type of impurities.

The scaling form applies provided that the system remains in the semiclassical metallic regime, the longitudinal resistivity is dominated by symmetric momentum-relaxing scattering, and terms beyond leading order in the antisymmetric scattering strength can be neglected.
\section{Conclusion and Discussion}
In summary, we have developed an angular-harmonic formulation of the semiclassical Boltzmann equation for nonlinear spin transport in the presence of skew scattering. Instead of replacing the collision integral by a single phenomenological relaxation time, we retain the angular and band dependence of the scattering kernel and decompose both the collision term and the nonequilibrium distribution function into Fourier harmonics. The resulting band-coupled algebraic equations can be solved recursively order by order in the applied electric field. Within the weak-skew-scattering approximation, the symmetric component of the distribution is obtained first, while the skew-induced correction is determined to leading order in the antisymmetric scattering rate.

We have applied this framework to the gapped surface states of a topological insulator with spin-dependent impurity scattering and evaluated the corresponding second-order extrinsic spin-current response. The results show that the nonlinear response is strongly influenced by the structure of the constant-energy contours and by the scattering channels available at the Fermi energy. In the regime $mv^{2}/\Delta>1$, the nonmonotonic lower-band dispersion supports two constant-energy contours over a finite energy interval, allowing scattering between the inner and outer contours. These additional processes are absent in the monotonic regime $mv^{2}/\Delta<1$, leading to qualitatively different chemical-potential dependences of the nonlinear spin-current coefficients. Pronounced response features occur when the chemical potential crosses characteristic band energies,where the constant-energy contours and the available scattering channels change rapidly.

These results demonstrate that higher angular harmonics of the nonequilibrium distribution and inter-contour scattering processes can be essential for describing extrinsic nonlinear spin transport. A single transport relaxation time, which primarily characterizes the leading angular harmonic, does not generally capture these effects. The present formulation can be extended to anisotropic Fermi surfaces, multiband systems, and more general impurity potentials. It therefore provides a systematic framework for investigating impurity-mediated nonlinear spin transport beyond the conventional relaxation-time approximation.

\begin{acknowledgments}
We thank Yuxuan Guo for insightful discussions and valuable perspectives that provided inspiration for this work.
\end{acknowledgments}

\appendix
\begin{widetext}
\section{\label{sec:apendixA}The calculation details of the electron distribution function}
When the parameters satisfy $mv^2>\Delta$, the energy band structure is similar to the energy band shown in Fig.~\ref{fig:band}(a), which has three energy regions. As outlined initially in the main results, the order-by-order Fourier expansion method of the electron distribution function provides a unified framework for treating all three regions, despite their distinct particle scattering patterns. \par 
The Boltzmann equantion in region $\mathrm{\uppercase\expandafter{\romannumeral 1}}$ has been shown in Eq.\ref{eq:14},\ref{eq:15},\ref{eq:16},\ref{eq:17}.  In region $\mathrm{\uppercase\expandafter{\romannumeral 1}}$, electrons can undergo transitions to states at the same energy within the same band, as well as to states at the same energy in a different band. The eigenstates and the T-matrix are shown in Eq.\ref{eq:10} and Eq.\ref{eq:13}.
We denote the momentum magnitude of an electron in the upper band as $k_{+}$, and that in the lower band as $k_{-}$, which satisfies:
\begin{equation}
    \frac{k^2}{2m}+h(k_{+})=\epsilon({k_{+}})=\epsilon(k_{-})=\frac{k_{-}^2}{2m}-h(k_{-})
\end{equation},where $h(k)=\sqrt{\Delta^2+k^2v^2}$.\\
The symmetric part of the transition probability is:
\begin{align}
    I_{++}(\theta_1-\theta_2)&=a^2\frac{(\Delta-h_{+})^4+k_{+}^4v^4+2(\Delta-h_{+})^2k_{+}^2v^2\cos(\theta_1-\theta_2)}{(k_{+}^2v^2+(\Delta-h_{+})^2)^2}\\
     I_{--}(\theta_1-\theta_2)&=a^2\frac{(\Delta+h_{-})^4+k_{-}^4v^4+2(\Delta+h_{-})^2k_{-}^2v^2\cos(\theta_1-\theta_2)}{(k_{-}^2v^2+(\Delta+h_{-})^2)^2}\\
     I_{+-}(\theta_1-\theta_2)&=a^2\frac{(\Delta-h_{+})^2(\Delta+h_{-})^2+k_{+}^2k_{-}^2v^4+2(\Delta-h_{+})(\Delta+h_{-})k_{+}k_{-}v^2\cos(\theta_1-\theta_2)}{(k_{+}^2v^2+(\Delta-h_{+})^2)(k_{-}^2v^2+(\Delta+h_{-})^2)}\\
     I_{-+}(\theta_1-\theta_2)&=I_{+-}(\theta_1-\theta_2)
\end{align}\\
The skew part of the transition probability is :
\begin{align}
    G_{++}(\theta_2-\theta_1)&=2ab\sin(\theta_2-\theta_1)\frac{k_{+}^2v^2-(\Delta-h_{+})^2}{k_{+}^2v^2+(\Delta-h_{+})^2}\\
    G_{--}(\theta_2-\theta_1)&=2ab\sin(\theta_2-\theta_1)\frac{k_{-}^2v^2-(\Delta+h_{-})^2}{k_{-}^2v^2+(\Delta+h_{-})^2}\\
    G_{+-}(\theta_2-\theta_1)&=2ab\sin(\theta_2-\theta_1)\frac{k_{+}^2k_{-}^2v^4-(\Delta-h_{+})^2(\Delta+h_{-})^2}{(k_{+}^2v^2+(\Delta-h_{+})^2)(k_{-}^2v^2+(\Delta+h_{-})^2)}\\
    G_{-+}(\theta_2-\theta_1)&=G_{+-}(\theta_2-\theta_1)
\end{align}\\
We can therefore acquire the Fourier expansion coefficients of $I(\theta)$ and $G(\theta)$.

We denote the transition matrix in Eq.~\ref{eq:16} and Eq.~\ref{eq:17}by:
\begin{equation}
    \mqty[\frac{m}{1+\frac{mv^2}{h_{+}}}(I_{++,l}-I_{++,0})-\frac{m}{1-\frac{mv^2}{h_{-}}}I_{+-,0}& \frac{m}{1-\frac{mv^2}{h_{-}}}I_{+-,l}\\
    \frac{m}{1+\frac{mv^2}{h_{+}}}I_{-+,l}&\frac{m}{1-\frac{mv^2}{h_{-}}}(I_{--,l}-I_{--,0})-\frac{m}{1+\frac{mv^2}{h_{+}}}I_{-+,0}]\equiv M_{l} 
\end{equation}
\begin{equation}
    \mqty[\frac{m}{1+\frac{mv^2}{h_{+}}}G_{++,l}& \frac{m}{1-\frac{mv^2}{h_{-}}}G_{+-,l}\\
    \frac{m}{1+\frac{mv^2}{h_{+}}}G_{-+,l}&\frac{m}{1-\frac{mv^2}{h_{-}}}G_{--,l}]\equiv iP_{l}
\end{equation}
Take n = 0, 
\begin{equation}
    \mqty[-e\vb{E}\cdot\pdv{f_{+,sym}^{(0)}(k_{+},\theta)}{\vb{k_{+}}}\\
    -e\vb{E}\cdot\pdv{f_{-,sym}^{(0)}(k_{-},\theta)}{\vb{k_{-}}}]
    =-eE\cos\theta\mqty[\dv{f^{(0)}_{+,sym}(k_{+})}{k_{+}}\\\dv{f^{(0)}_{-,sym}(k_{-})}{k_{-}}]
    \label{ap:eq1}
\end{equation}
Therefore, the Fourier expansion coefficients of the \ref{ap:eq1} are nonzero only when $l=\pm1$. The Eq.\ref{eq:16} with $l=1$ can be written as:
\begin{equation}
    -\frac{eE}{2}\mqty[\dv{f^{(0)}_{+,sym}(k_{+})}{k_{+}}\\\dv{f^{(0)}_{-,sym}(k_{-})}{k_{-}}]= M_{1}\mqty[f_{+,sym,1}^{(1)}(k_{+})\\ f_{-,sym,1}^{(1)}(k_{-})]
\end{equation}
If we denote the inverse of $M_l$ as $Q_l$,we obtain :
\begin{align}
    \mqty[f_{+,sym,1}^{(1)}(k_{+})\\ f_{-,sym,1}^{(1)}(k_{-})]&=-\frac{eE}{2}Q_{1}\mqty[\dv{f^{(0)}_{+,sym}(k_{+})}{k_{+}}\\\dv{f^{(0)}_{-,sym}(k_{-})}{k_{-}}]\\
    \mqty[f_{+,sym}^{(1)}(k_{+},\theta)\\ f_{-,sym}^{(1)}(k_{-},\theta)]&=-eEQ_{1}\mqty[\dv{f^{(0)}_{+,sym}(k_{+})}{k_{+}}\\\dv{f^{(0)}_{-,sym}(k_{-})}{k_{-}}]\cos\theta
\end{align}
In polar coordinates,$\partial_{\vb{k}}=\hat{k}\partial_{k}+\frac{1}{k}\hat{\theta}\partial_{\theta}$, we can get the Eq.\ref{eq:16} with $n=1$:
\begin{equation}
    -e\vec{E}\cdot\partial_{\vb{k}}(-eE\cos\theta Q_{1}\mqty[\dv{f^{(0)}_{+,sym}(k_{+})}{k_{+}}\\\dv{f^{(0)}_{-,sym}(k_{-})}{k_{-}}])=e^2E^2\cos^2\theta\mqty[\dv{k_{+}}&0\\0&\dv{k_{-}}]Q_{1}\mqty[\dv{f^{(0)}_{+,sym}(k_{+})}{k_{+}}\\\dv{f^{(0)}_{-,sym}(k_{-})}{k_{-}}]+e^2E^2\sin^2\theta \mqty[\frac{1}{k_{+}}&0\\0&\frac{1}{k_{-}}]Q_1\mqty[\dv{f^{(0)}_{+,sym}(k_{+})}{k_{+}}\\\dv{f^{(0)}_{-,sym}(k_{-})}{k_{-}}]
    \label{ap:eq2}
\end{equation}
We focus exclusively on the non-zero frequency components of the electron distribution function, as the zero-frequency component does not contribute to the spin current transport. 
The $l=2$ part of the Eq.\ref{ap:eq2} is:
\begin{equation}
    \frac{e^2E^2}{4}\mqty[\dv{k_{+}}-\frac{1}{k_{+}}&0\\0&\dv{k_{-}}-\frac{1}{k_{-}}]Q_{1}\mqty[\dv{f^{(0)}_{+,sym}(k_{+})}{k_{+}}\\\dv{f^{(0)}_{-,sym}(k_{-})}{k_{-}}]
\end{equation}
Using the $l=2,n=1$ part of Eq.\ref{eq:16}, we can acquire:
\begin{equation}
    \mqty[f_{+,sym}^{(2)}(k_{+},\theta)\\ f_{-,sym}^{(2)}(k_{-},\theta)]=\frac{e^2E^2}{2}\cos (2\theta) Q_{2}\mqty[\dv{k_{+}}-\frac{1}{k_{+}}&0\\0&\dv{k_{-}}-\frac{1}{k_{-}}]Q_{1}\mqty[\dv{f^{(0)}_{+,sym}(k_{+})}{k_{+}}\\\dv{f^{(0)}_{-,sym}(k_{-})}{k_{-}}]
\end{equation}
We can acquire the skew part of the electron distribution from the Eq.~\ref{eq:17}:
\begin{equation}
    0=iP_{1}\mqty[f_{+,sym,1}^{(1)}(k_{+})\\f_{-,sym,1}^{(1)}(k_{-})]+M_{1}\mqty[f_{+,sk,1}^{(1)}(k_{+})\\f_{-,sk,1}^{(1)}(k_{-})]\Rightarrow 
    \mqty[f_{+,sk,1}^{(1)}(k_{+})\\f_{-,sk,1}^{(1)}(k_{-})]=-iQ_{1}P_{1}\mqty[f_{+,sym,1}^{(1)}(k_{+})\\f_{-,sym,1}^{(1)}(k_{-})]
\end{equation}
\begin{equation}
    \mqty[f_{+,sk}^{(1)}(k_{+},\theta)\\f_{-,sk}^{(1)}(k_{-},\theta)]=2\sin\theta Q_{1}P_{1}\mqty[f_{+,sym,1}^{(1)}\\f_{-,sym,1}^{(1)}]
\end{equation}
\begin{equation}
    -e\vb{E}\cdot\mqty[\pdv{f_{+,sk}^{(1)}(k_{+},\theta)}{k_{+}}\\\pdv{f_{-,sk}^{(1)}(k_{-},\theta)}{k_{-}}]=ieE\sin(2\theta)\mqty[\frac{1}{k_{+}}-\dv{k_{+}}&0\\0&\frac{1}{k_{-}}-\dv{k_{-}}]\mqty[f_{+,sk,1}^{(1)}\\f_{-,sk,1}^{(1)}]
\end{equation}
According to the Eq\ref{eq:17} at $l=2,n=1$:
we can acquire the second-order skew part of the electron distribution :
\begin{align}
    \mqty[f_{+,sk}^{(2)}(k_{+},\theta)\\f_{-,sk}^{(2)}(k_{-},\theta)]&=ieE\sin(2\theta)Q_2\mqty[\frac{1}{k_{+}}-\dv{k_{+}}&0\\0&\frac{1}{k_{-}}-\dv{k_{-}}]\mqty[f_{+,sk,1}^{(1)}\\f_{-,sk,1}^{(1)}]\\
    &= eE\sin(2\theta)Q_2\mqty[\frac{1}{k_{+}}-\dv{k_{+}}&0\\0&\frac{1}{k_{-}}-\dv{k_{-}}]Q_{1}P_{1}\mqty[f_{+,sym,1}^{(1)}\\f_{-,sym,1}^{(1)}]
\end{align},where we use the property that $G(\theta)$ has no the $l=0$ and $l=2$ Fourier coefficients.
The electron distribution function in region $\mathrm{\uppercase\expandafter{\romannumeral 1}}$ has been calculated. In region $\mathrm{\uppercase\expandafter{\romannumeral 2}}$, the collision item is relatively simpler because electron transitions are only permitted exclusively within the constant-energy contour of the lower band:
\begin{align}
    [-e\vb{E}\cdot\partial_{k_{-}}f_{-,sym}^{(n)}]_{l}&=\frac{m}{1-\frac{mv^2}{h_{-}}}(I_{--,l}-I_{--,0})f_{-,sym,l}^{(n+1)}\\
    [-e\vb{E}\cdot\partial_{k_{-}}f_{-,sk}^{(n)}]_{l}&=iP_{--,l}\frac{m}{1-\frac{mv^2}{h_{-}}}f_{-,sym,l}^{(n+1)}+(I_{--,l}-I_{--,0})\frac{m}{1-\frac{mv^2}{h_{-}}}f_{-,sk,l}^{(n+1)}
\end{align}
Similar to the calculation in region $\mathrm{\uppercase\expandafter{\romannumeral 1}}$,we have:
\begin{align}
f_{-,sym}^{(1)}(k_{-},\theta)&=-eE\cos\theta \frac{\dv{f_{-,sym}^{(0)}(k_{-})}{k_{-}}}{\frac{m}{1-\frac{mv^2}{h_{-}}}(I_{--,1}-I_{--,0})}\\
f_{-,sk}^{(1)}(k_{-},\theta)&= 2\sin\theta \frac{P_{--,1}}{I_{--,1}-I_{--,0}}f_{-,sym,1}^{(1)}(k_{-})\\
f_{-,sym}^{(2)}(k_{-},\theta)&=eE\cos(2\theta)\frac{(\dv{k_{-}}-\frac{1}{k_{-}})}{\frac{m}{1-\frac{mv^2}{h_{-}}}I_{--,0}}f_{-,sym,1}^{(1)}(k_{-})\\
f_{-,sk}^{(2)}(k_{-},\theta)&=ieE\sin(2\theta)\frac{(\dv{k_{-}}-\frac{1}{k_{-}})}{\frac{m}{1-\frac{mv^2}{h_{-}}}I_{--,0}}f_{-,sk,1}^{(1)}(k_{-})
\end{align}
In region $\mathrm{\uppercase\expandafter{\romannumeral 3}}$, for a given energy, there exist two momentum solutions with different magnitudes: $k_{+}$ and $k_{-}$, where $k_{+}$ represents the magnitude of momentum of inner contour and $k_{-}$ represents the magnitude of momentum of outer circle. Electron transitions can occur between the inner and outer contours. Since the definition of the $+$ and $-$ does not represent the upper and lower band, we need to deduce the Boltzmann Equation under new situation:
\begin{align}
    -e\vb{E}\cdot\pdv{f_{+,sym}^{(n)}(k_{+},\theta)}{\vb{k_{+}}}&=\iint \frac{k'dk'd\theta'}{(2\pi)^2}2\pi\delta(\epsilon_k-\epsilon_{k'})[C_{+-}(\theta-\theta')f_{-,sym}^{(n+1)}(k',\theta')-C_{-+}(\theta-\theta')f_{+,sym}^{(n+1)}(k_{+},\theta)\nonumber\\ &+C_{++}(\theta-\theta')f_{+,sym}^{(n+1)}(k',\theta')-C_{++}(\theta-\theta')f_{+,sym}^{(n+1)}(k_{+},\theta)]\nonumber \\
    &=\int_{0}^{2\pi}\frac{d\theta'}{2\pi}(\frac{m}{1-\frac{mv^2}{h_{-}}}(C_{+-}(\theta-\theta')f_{-,sym}^{(n+1)}(k_{-},\theta')-C_{-+}(\theta-\theta')f_{+,sym}^{(n+1)}(k_{+},\theta)\nonumber\\&+\frac{m}{\frac{mv^2}{h_{+}}-1}(C_{++}(\theta-\theta')f_{+,sym}^{(n+1)}(k_{+},\theta')-C_{++}(\theta-\theta')f_{+,sym}^{(n+1)}(k_{+},\theta))
    \label{ap:eq3}
\end{align}
\begin{align}
          -e\vb{E}\cdot\pdv{f_{-,sym}^{(n)}(k_{-},\theta)}{\vb{k_{-}}}&=\iint \frac{k'dk'd\theta'}{(2\pi)^2}2\pi\delta(\epsilon_k-\epsilon_{k'})[C_{-+}(\theta-\theta')f_{+,sym}^{(n+1)}(k',\theta')-C_{+-}(\theta-\theta')f_{-,sym}^{(n+1)}(k_{-},\theta)\nonumber\\ &+C_{--}(\theta-\theta')f_{-,sym}^{(n+1)}(k',\theta')-C_{--}(\theta-\theta')f_{-,sym}^{(n+1)}(k_{-},\theta)]\nonumber \\
    &=\int_{0}^{2\pi}\frac{d\theta'}{2\pi}(\frac{m}{\frac{mv^2}{h_{+}}-1}(C_{-+}(\theta-\theta')f_{+,sym}^{(n+1)}(k_{+},\theta')-C_{+-}(\theta-\theta')f_{-,sym}^{(n+1)}(k_{-},\theta)\nonumber\\&+\frac{m}{1-\frac{mv^2}{h_{-}}}(C_{--}(\theta-\theta')f_{-,sym}^{(n+1)}(k_{-},\theta')-C_{--}(\theta-\theta')f_{-,sym}^{(n+1)}(k_{-},\theta))
    \label{ap:eq4}
\end{align}
To distinguish from the notation in region $\mathrm{\uppercase\expandafter{\romannumeral1}}$, we use $C(\theta)$ to represent the symmetric part of the scattering probability and $D(\theta)$ to represent the skew part of the scattering probability. Similarly to the formula in region $\mathrm{\uppercase\expandafter{\romannumeral1}}$,the Fourier expansion of Eq.~\ref{ap:eq3} and Eq.~\ref{ap:eq4} can be written in matrix forms:
\begin{equation}
    \mqty[\frac{m}{-1+\frac{mv^2}{h_{+}}}(C_{++,l}-C_{++,0})-\frac{m}{1-\frac{mv^2}{h_{-}}}C_{+-,0}& \frac{m}{1-\frac{mv^2}{h_{-}}}C_{+-,l}\\
    \frac{m}{-1+\frac{mv^2}{h_{+}}}C_{-+,l}&\frac{m}{1-\frac{mv^2}{h_{-}}}(C_{--,l}-C_{--,0})-\frac{m}{-1+\frac{mv^2}{h_{+}}}C_{-+,0}]
    \mqty[f_{+,sym,l}^{(n+1)}(k_{+})\\f_{-,sym,l}^{(n+1)}(k_{-})]=\mqty[[-e\vb{E}\cdot\pdv{f_{+,sym}^{(n)}(k_{+},\theta)}{\vb{k_{+}}}]_{l}\\
    [-e\vb{E}\cdot\pdv{f_{-,sym}^{(n)}(k_{-},\theta)}{\vb{k_{-}}}]_{l}
    ]
    \label{ap:eq4a}
\end{equation}
The skew part of the Boltzmann equation can be written as:
\begin{align}
    \mqty[[-e\vb{E}\cdot\pdv{f_{+,sk}^{(n)}(k_{+},\theta)}{\vb{k_{+}}}]_{l}\\
    [-e\vb{E}\cdot\pdv{f_{-,sk}^{(n)}(k_{-},\theta)}{\vb{k_{-}}}]_{l}
    ] &=
        \mqty[\frac{m}{-1+\frac{mv^2}{h_{+}}}(C_{++,l}-C_{++,0})-\frac{m}{1-\frac{mv^2}{h_{-}}}C_{+-,0}& \frac{m}{1-\frac{mv^2}{h_{-}}}C_{+-,l}\\
    \frac{m}{-1+\frac{mv^2}{h_{+}}}C_{-+,l}&\frac{m}{1-\frac{mv^2}{h_{-}}}(C_{--,l}-C_{--,0})-\frac{m}{-1+\frac{mv^2}{h_{+}}}C_{-+,0}]
    \mqty[f_{+,sk,l}^{(n+1)}(k_{+})\\f_{-,sk,l}^{(n+1)}(k_{-})]\nonumber\\
    &+\mqty[\frac{m}{-1+\frac{mv^2}{h_{+}}}D_{++,l}& \frac{m}{1-\frac{mv^2}{h_{-}}}D_{+-,l}\\
    \frac{m}{-1+\frac{mv^2}{h_{+}}}D_{-+,l}&\frac{m}{1-\frac{mv^2}{h_{-}}}D_{--,l}]\mqty[f_{+,sym,l}^{(n+1)}(k_{+})\\f_{-,sym,l}^{(n+1)}(k_{-})]
    \label{ap:eq5}
\end{align}
The symmetric part of the transition probability is:
\begin{align}
     C_{++}(\theta_1-\theta_2)&=a^2\frac{(\Delta+h_{+})^4+k_{+}^4v^4+2(\Delta+h_{+})^2k_{+}^2v^2\cos(\theta_1-\theta_2)}{(k_{+}^2v^2+(\Delta+h_{+})^2)^2}\\
     C_{--}(\theta_1-\theta_2)&=a^2\frac{(\Delta+h_{-})^4+k_{-}^4v^4+2(\Delta+h_{-})^2k_{-}^2v^2\cos(\theta_1-\theta_2)}{(k_{-}^2v^2+(\Delta+h_{-})^2)^2}\\
     C_{+-}(\theta_1-\theta_2)&=a^2\frac{(\Delta+h_{+})^2(\Delta+h_{-})^2+k_{+}^2k_{-}^2v^4+2(\Delta+h_{+})(\Delta+h_{-})k_{+}k_{-}v^2\cos(\theta_1-\theta_2)}{(k_{+}^2v^2+(\Delta+h_{+})^2)(k_{-}^2v^2+(\Delta+h_{-})^2)}\\
     C_{-+}(\theta_1-\theta_2)&=C_{+-}(\theta_1-\theta_2)
\end{align}
The skew part of the transition probability is:
\begin{align}
      D_{++}(\theta_2-\theta_1)&=2ab\sin(\theta_2-\theta_1)\frac{k_{+}^2v^2-(\Delta+h_{+})^2}{k_{+}^2v^2+(\Delta+h_{+})^2}\\
    D_{--}(\theta_2-\theta_1)&=2ab\sin(\theta_2-\theta_1)\frac{k_{-}^2v^2-(\Delta+h_{-})^2}{k_{-}^2v^2+(\Delta+h_{-})^2}\\
    D_{+-}(\theta_2-\theta_1)&=2ab\sin(\theta_2-\theta_1)\frac{k_{+}^2k_{-}^2v^4-(\Delta+h_{+})^2(\Delta+h_{-})^2}{(k_{+}^2v^2+(\Delta+h_{+})^2)(k_{-}^2v^2+(\Delta+h_{-})^2)}\\
    D_{-+}(\theta_2-\theta_1)&=D_{+-}(\theta_2-\theta_1)
\end{align}
Since the structure of the Eq.\ref{ap:eq4} and \ref{ap:eq5} is the same as the Eq.\ref{eq:16} and \ref{eq:17}, If we denote the transition matrix by:
\begin{equation}
    \mqty[\frac{m}{-1+\frac{mv^2}{h_{+}}}(C_{++,l}-C_{++,0})-\frac{m}{1-\frac{mv^2}{h_{-}}}C_{+-,0}& \frac{m}{1-\frac{mv^2}{h_{-}}}C_{+-,l}\\
    \frac{m}{-1+\frac{mv^2}{h_{+}}}C_{-+,l}&\frac{m}{1-\frac{mv^2}{h_{-}}}(C_{--,l}-C_{--,0})-\frac{m}{-1+\frac{mv^2}{h_{+}}}C_{-+,0}]
   \equiv M^{\prime}_{l}
\end{equation}
\begin{equation}
    \mqty[\frac{m}{-1+\frac{mv^2}{h_{+}}}D_{++,l}& \frac{m}{1-\frac{mv^2}{h_{-}}}D_{+-,l}\\
    \frac{m}{-1+\frac{mv^2}{h_{+}}}D_{-+,l}&\frac{m}{1-\frac{mv^2}{h_{-}}}D_{--,l}]\equiv iP^{\prime}_{l}
\end{equation}
We can just substitute $M_{l},Q_{l},P_{l}$ with $M^{\prime}_{l},Q^{\prime}_{l},P^{\prime}_{l}$ to acquire the electron distribution in region $\mathrm{\uppercase\expandafter{\romannumeral3}}$.
It is important to note that when differentiating with respect to $k_{+}$, $k_{-}$ must be treated as a function of $k_{+}$, and vice versa.

\section{\label{sec:appendixB}The second-order spin current in the model Hamiltonian}
Due to the rotational symmetry of the system about the z-axis, we can always choose the electric field to be applied along the x-direction. Therefore, we just need to calculate $\Omega^{m}_{nx},j^{m}_{n}$:\\
For the upper energy band:
\begin{align}
    j^{x}_{x,+}&=\mel{u_{+}}{\frac{1}{4}\{\pdv{H}{k_{x}},\sigma^x\}}{u_{+}}=\frac{k_{x}}{2m}\mel{u_{+}}{\sigma^{x}}{u_{+}}=\frac{k^2v(\Delta-h)\sin\theta\cos\theta}{m(k^2v^2+(\Delta-h)^2)}\\
    j^{y}_{x,+}&=\mel{u_{+}}{\frac{1}{4}\{\pdv{H}{k_{x}},\sigma^y\}}{u_{+}}=\mel{u_{+}}{\frac{1}{2}v+\frac{k_x}{2m}\sigma^{y}}{u_{+}}=\frac{v}{2}-\frac{k^2v(\Delta-h)\cos^2\theta}{m(k^2v^2+(\Delta-h)^2)}\\
    j^{x}_{y,+}&=\mel{u_{+}}{\frac{1}{4}\{\pdv{H}{k_{y}},\sigma^x\}}{u_{+}}=\mel{u_{+}}{-\frac{1}{2}v+\frac{k_y}{2m}\sigma^{x}}{u_{+}}=-\frac{v}{2}+\frac{k^2v(\Delta-h)\sin^2\theta}{m(k^2v^2+(\Delta-h)^2)}\\
    j^{y}_{y,+}&=\mel{u_{+}}{\frac{1}{4}\{\pdv{H}{k_{y}},\sigma^y\}}{u_{+}}=\frac{k_{y}}{2m}\mel{u_{+}}{\sigma^{y}}{u_{+}}=-\frac{k^2v(\Delta-h)\sin\theta\cos\theta}{m(k^2v^2+(\Delta-h)^2)}\\  
\end{align}
For the lower energy band:
\begin{align}
    j^{x}_{x,-}&=\mel{u_{-}}{\frac{1}{4}\{\pdv{H}{k_{x}},\sigma^x\}}{u_{-}}=\frac{k_{x}}{2m}\mel{u_{-}}{\sigma^{x}}{u_{-}}=\frac{k^2v(\Delta+h)\sin\theta\cos\theta}{m(k^2v^2+(\Delta+h)^2)}\\
    j^{y}_{x,-}&=\mel{u_{-}}{\frac{1}{4}\{\pdv{H}{k_{x}},\sigma^y\}}{u_{-}}=\mel{u_{-}}{\frac{1}{2}v+\frac{k_x}{2m}\sigma^{y}}{u_{-}}=\frac{v}{2}-\frac{k^2v(\Delta+h)\cos^2\theta}{m(k^2v^2+(\Delta+h)^2)}\\
    j^{x}_{y,-}&=\mel{u_{-}}{\frac{1}{4}\{\pdv{H}{k_{y}},\sigma^x\}}{u_{-}}=\mel{u_{-}}{-\frac{1}{2}v+\frac{k_y}{2m}\sigma^{x}}{u_{-}}=-\frac{v}{2}+\frac{k^2v(\Delta+h)\sin^2\theta}{m(k^2v^2+(\Delta+h)^2)}\\
    j^{y}_{y,-}&=\mel{u_{-}}{\frac{1}{4}\{\pdv{H}{k_{y}},\sigma^y\}}{u_{-}}=\frac{k_{y}}{2m}\mel{u_{-}}{\sigma^{y}}{u_{-}}=-\frac{k^2v(\Delta+h)\sin\theta\cos\theta}{m(k^2v^2+(\Delta+h)^2)}\\  
\end{align}
According to the Eq.\ref{eq:18}:
\begin{align}
    \Omega^{x}_{xx,+}&=-\Omega^{x}_{xx,-}=\frac{k^3v^3\Delta \cos\theta}{m(k^2v^2+(\Delta-h)^2)(k^2v^2+(\Delta+h)^2)h}\\
    \Omega^{y}_{xx,+}&=-\Omega^{y}_{xx,-}=0\\
    \Omega^{x}_{yx,+}&=-\Omega^{x}_{yx,-}=\frac{k^3v^3\Delta \sin\theta}{m(k^2v^2+(\Delta-h)^2)(k^2v^2+(\Delta+h)^2)h}\\
    \Omega^{y}_{yx,+}&=-\Omega^{y}_{yx,-}=0
\end{align}

A limitation of the present formulation concerns the zeroth angular harmonic of the second-order nonequilibrium distribution. Because the impurity scattering considered here is purely elastic, it only redistributes electrons among states on the same constant-energy contour and therefore cannot relax an isotropic energy-dependent distortion of the distribution function. Consequently, the collision operator possesses a zero mode in the $l=0$ sector, and the corresponding collision matrix is singular. This issue first becomes relevant at second order in the electric field, since the drift term acting on the first-order $l=1$ distribution generates both $l=0$ and $l=2$ components. The $l=2$ component describes an angular anisotropy and is fully relaxed by elastic impurity scattering, whereas the $l=0$ component is associated with an isotropic redistribution of electronic energy, closely related to Joule heating, and requires an inelastic energy-relaxation mechanism to reach a steady state. A complete treatment of this contribution would require additional inelastic processes, such as electron-phonon or other energy-relaxing scattering mechanisms. In the present work, we therefore restrict our calculation to the anisotropic nonequilibrium components that are uniquely determined by the elastic impurity collision operator and neglect the undetermined $l=0$ second-order contribution.
\end{widetext}

\bibliography{apssamp}
\end{document}